\documentclass[conference]{IEEEtran}
\IEEEoverridecommandlockouts
\usepackage{cite}
\usepackage{amsmath,amssymb,amsfonts}
\usepackage{algorithmic}
\usepackage{graphicx}
\usepackage{textcomp}
\def\BibTeX{{\rm B\kern-.05em{\sc i\kern-.025em b}\kern-.08em
		T\kern-.1667em\lower.7ex\hbox{E}\kern-.125emX}}
\begin{document}
	
	\title{Big Data Analytic based on Scalable PANFIS for RFID Localization\\}
	
	\author{\IEEEauthorblockN{1\textsuperscript{st} Choiru Za'in}
		\IEEEauthorblockA{\textit{Department of Computer Science and Information Technology} \\
			\textit{La Trobe University}\\
			Melbourne, Australia\\
			C.Zain@latrobe.edu.au}
		\and
		\IEEEauthorblockN{2\textsuperscript{nd} Mahardhika Pratama}
		\IEEEauthorblockA{\textit{School of Computer Science and Engineering} \\
			\textit{Nanyang Technological University}\\
			Singapore, Singapore\\
			mpratama@ntu.edu.sg}
		\and
		\IEEEauthorblockN{3\textsuperscript{rd} Andri Ashfahani}
		\IEEEauthorblockA{\textit{School of Computer Science and Engineering} \\
			\textit{Nanyang Technological University}\\
			Singapore, Singapore\\
			andriash@e.ntu.edu.sg}
		\and
		\IEEEauthorblockN{4\textsuperscript{th} Eric Pardede}
		\IEEEauthorblockA{\textit{Department of Computer Science and Information Technology} \\
			\textit{La Trobe University}\\
			Melbourne, Australia\\
			E.Pardede@latrobe.edu.au}
		\and
		\IEEEauthorblockN{5\textsuperscript{th} Huang Sheng}
		\IEEEauthorblockA{\textit{} \\
			\textit{Singapore Institute of Manufacturing Technology}\\
			Singapore, Singapore\\
			shuang@SIMTech.a-star.edu.au}
		\\
	}
	
	\maketitle
	
	\begin{abstract}
		
		RFID technology has gained popularity to address localization problem in the manufacturing shopfloor due to its affordability and easiness in deployment. This technology is used to track the manufacturing object location to increase the production's efficiency. However, the data used for localization task is not easy to analyze because it is generated from the non-stationary environment. It also continuously arrive over time and yields the large-volume of data. Therefore, an advanced big data analytic is required to overcome this problem. 
		We propose a distributed big data analytic framework based on PANFIS (Scalable PANFIS), where PANFIS is an evolving algorithm  which has capability to learn data stream in the single pass mode. Scalable PANFIS can learn big data stream by processing many chunks/partitions of data stream. Scalable PANFIS is also equipped with rule' structure merging to eliminate the redundancy among rules. Scalable PANFIS is validated by measuring its performance against single PANFIS and other Spark's scalable machine learning algorithms. The result shows that Scalable PANFIS performs running time more than 20 times faster than single PANFIS. The rule merging process in Scalable PANFIS shows that there is no significant reduction of accuracy in classification task with 96.67 percent of accuracy in comparison with single PANFIS of 98.71 percent. Scalable PANFIS also generally outperforms some Spark MLib machine learnings to classify RFID data with the comparable speed in running time.
	\end{abstract}
	
	
	\section{Introduction}
	
	Radio Frequency Identification (RFID) localization systems (RFID localization) has become a popular technology in manufacturing or other production base industries to optimize the production output by providing the position/location of the Manufacturing Objects (MOs) (e.g. workstations, tools, staffs, and  support services). In the manufacturing process, obtaining the accurate location of the MOs is essential because this information can represent the current Work-In-Process (WIP) condition in the industry. The WIP becomes the basis reference for decision making to avoid unexpected failure and in turn will increase the production output.

	RFID localization technology is considered to be chosen among  other localization technologies such as Wireless Sensor Networks (WSNs), Ultrasound, Infrared, and Video Camera due to its simplicity and price affordability in the deployment process \cite{ni2011rfid}. These benefits allow industry to apply RFID localization devices easily in many areas of the manufacturing location in order to provide the accurate MOs to the Manufacturing Execution Systems (MES).

	RFID localization systems consist of some devices and components: RFID tag, RFID reader, and data processing subsystems. RFID tag transmits beacon messages to RFID reader. RFID reader captures the Received Signal Strength (RSS) along with the tag ID from RFID tag. These signals arrive continuously and they need to be processed by the MES in the real-time mode.
	
	Processing and analyzing MOs locations from the received RSS in RFID localization are challenging tasks due to two factors: 1) RSS information are not always reliable because its value varies due to environmental
	changes (e.g. minor movement); 2) RSS information arrives in the real-time mode rapidly, which can be considered as a data stream. Thus, it will cause the generation of large-volume of data which is stored in the cloud/server, and this phenomenon can be considered as a big data problem. All of these factors underlie the needs of the advanced analytic algorithms to discover the knowledge of a big data stream.
	
	It has been a common practice that machine learning algorithms are used for knowledge discovery of the data for decision making purposes. Many new business requirements are growing based on big data stream analytic services. However, discovering big data stream knowledge for decision making is challenging due to its 4V\textquoteright s characteristics: volume, velocity, variety, and veracity. For this reason, in RFID localization task, an advanced big data analytic is required to enable faster and accurate decisions in terms of determining the MOs locations to catch up the expected production speed.
	
	Evolving machine learning techniques have gained popularity to process data stream, such as the work conducted in \cite{za2017evolving} in web news mining classification. These techniques aim to learn non-stationary data by learning the data in the single pass mode, enable to update data pattern for every incoming datum, without retraining all history data \cite{pratama2016evolving,pratama2015pclass,lughofer2015generalized,sayed2012learning}. This feature is beneficial in handling data stream to cope with the velocity characteristic of big data. Furthermore, there are many distributed platforms such as Hadoop \cite{shvachko2010hadoop} and Spark \cite{zaharia2010spark} to scale the limited resource of CPU in processing and storing big data. With many challenging issues in big data, most of the research directions in this area are growing towards a development of the advanced distributed big data analytic to discover big data stream knowledge.
	
	Due to the increasing demand in big data knowledge discovery, many efforts have been made on building large-scale distributed machine learning. Some recent works have been conducted in \cite{fernandez2014big,zabig}.
	Both frameworks incorporate distributed processing
	platform to scale the machine learning capability by dividing data stream
	into many chunks. However, while the first only processes the chunk of big data
	in batch mode as a standard platform for distributed learning, the latter extends the big data analytic framework to address the velocity characteristic of big data, enabling data chunk to be processed as a data stream, using evolving algorithm namely PANFIS \cite{pratama2014panfis}. The main property of the work in \cite{zabig} is the model fusion, where all models generated from all data chunks are merged into a final model. This final model represents the current knowledge big data.
	
	Based on the previous work in \cite{zabig}, we propose a novel big data stream analytic based on PANFIS, a seminal evolving neuro fuzzy system,
	to classify RFID big data stream for localization task. The distributed big data stream in RFID localization analytic framework has the following characteristics:

	\begin{enumerate}
		\item It processes the rapid rate of RSS signals using 
		big data analytic framework by incorporating PANFIS, where PANFIS processes data stream in the single pass mode, and distributed machine learning framework using Spark platform.
		\item This framework is equipped with rule' structure merging and rule parameters update to cope with the changing environment. 
	\end{enumerate}
	
	The rest of the paper is organized as follows. Section 2 discusses
	the related researches: RFID localization systems and a technique to capture the RFID data from sensors. Section 3 describes our proposed approach which
	specifically describes the data streams' flow processing in Spark platform. Section 4 discusses experimental setup and results in evaluating
	big data analytic and section 5 will conclude the paper.
	
	\section{RFID Localization Problems}
	
	\subsection{RFID Localization Systems\label{subsec:rfidapps}}

	On the manufacturing shopfloor, there are many MOs required to be
	stored, received, installed, and tracked. The conventional approach
	to manage MOs drawn a lot of human resources. Thus, advanced techniques
	are required to identify and track MOs location in the large manufacturing
	shopfloor. Radio Frequency Identification (RFID) localization technology is one of many technologies which has been put forward for localization system and tracking due to its price affordability and the easiness of deployment.
	
	The RFID localization system consists of three main components: RFID tag, RFID reader, and data processing subsystem. The main functionality of an RFID tag is to send the beacon message containing the tag ID and radio signal strength (RSS) information. The difference between active and passive tags lies on its power. RFID tag is categorized into two types: active tags and passive tags. Active tag is battery powered, which actively sending the beacon message periodically to the RFID reader, whereas the passive tag does not have a power source. Thus, the signal range of active tag can reached up to 300 m compared to 1 m for a passive tag. This experiment utilizes a passive tag, which only sends a beacon message once it receives the signal from the RFID reader. 
	
	RFID reader, as a second component in RFID localization system, captures the signals (data) sent by the tags and transmits it to the data processing subsystem. These signals are then processed by using the algorithm embedded in the data processing subsystems to do the RFID localization task. As a result, data processing subsystem provides the accurate MOs in the manufacturing shopfloor.
	
	The most difficult task in the RSS processing task is the noise attached in the RSS. In an ideal condition, MOs location can be determined using the normal measured RSS value. However, due to the severe interference and multi-path effect, the
	RSS information becomes unreliable and keeps changing overtime. A
	minor change in the environment can cause a significant change in the
	RSS information. Thus, only calculating distances from RSS observations does not solve the localization problem. Therefore, an advanced machine learning algorithm is employed to train the value of RSS based on reference tags information to accurately perform the localization task.
	The experiment is set up by deploying the reference tags on the fixed and known locations. The RSS information from reference tags are exploited to assist in locating
	the MOs. Suppose there is an RFID network which consists of $n$ tags
	$T=\{T_{i}:i=1,\dots,n\}$ and $m$ readers $R=\{R_{j}:j=1,\dots,m\}$.
	The RSS observations of tag $T_{i}$ at time-step $t$ is given as
	$x_{i}(t)=[x_{i1},\dots,x_{im}]^{T}$, where $x_{im}$ is the RSS
	observation of the tag $T_{i}$ by reader $R_{m}$. The position of
	reference tag is denoted as class label $y(t)=[y_{1},\dots,y_{n}]$,
	and thus the localization can be formulated into a multi-class classification
	problem. The Scalable PANFIS embedded in data processing subsystem is then employed as the machine learning algorithm to determine the MOs locations in the real-time mode.
	\subsection{Capturing RFID Data From Sensor}
\begin{figure}[htbp]
		\begin{centering}
			\includegraphics[width=9cm]{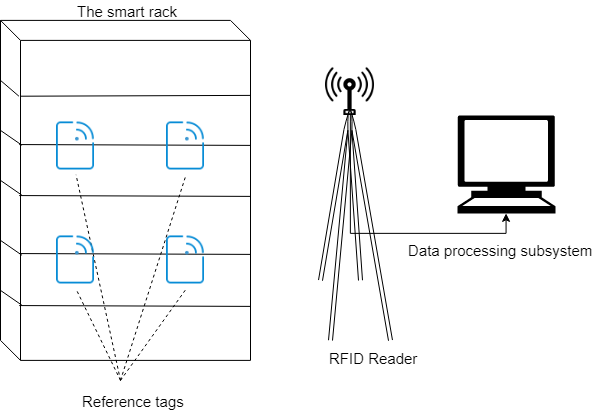}
			\par\end{centering}
		\caption{Data collection procedure in RFID Localization Systems \label{Figureandri}}
	\end{figure}

	This localization setup were conducted at SIMTech laboratory, Singapore. The environment is arranged to represents the RFID smart rack system. The system aims to improve the production efficiency in manufacturing process by locating the right material for production. Thus, the accurate information of MOs location is highly required. As can be seen in Fig. \ref{Figureandri}, one RFID reader and four RFID reference tags along with data processing subsystem are set up. The objects are placed at a steel storage rack with the dimensions of $1510 \thickspace mm \times 600 \thickspace mm \times 2020 \thickspace mm$. The rack has 5 different shelves, each of them $360 \thickspace mm$ apart. The rack shelf can contain up to 6 test objects, and each of them is attached with a passive RFID tag. The antenna is installed close to the rack at a distance of $1000 \thickspace mm$ from the rack. The height of the antenna is $2200\thickspace mm$ above the ground and it is connected to the RFID receiver. The receiver is then connected to a data processing subsystem, where the localization algorithm is executed, through the ethernet links. The number of reference tag indicates that there are 4 classes representing the location of the objects considered in this experiments.
	
	The data processing subsystem has two main components: data acquisition component and algorithm execution component. The Microsoft Visual C++ based PC application is developed to acquire the RSS information from all tags data, while the localization algorithm is executed using big data analytic based on PANFIS. The  reader is also configured to record the RSS information for every $1\thickspace s$. During  $20 \thickspace hours$ of experiment recording, the number of 283,100 observations are obtained, where every instance has the information of RSS and its reference tag which represents the MOs object location. Based on the recorded RFID data, Scalable PANFIS can learn this data stream behavior to predict the future data.
	\section{Distributed Big Data Analytic Based on PANFIS Approach}
	This section presents the details of our distributed big data
	analytic based on PANFIS. The briefly overview of PANFIS algorithm is described in the subsection \ref{subsec:PANFISALG}. Big data analytic architecture and data processing are elaborated in the subsection \ref{subsec:BigDataFlow}. Furthermore, subsection \ref{subsec:RuleMerging} will describe the merging process of all rules generated from PANFIS learning algorithm of every data chunk.
	\subsection{PANFIS Algorithm\label{subsec:PANFISALG}}
	
	PANFIS is built based on Neuro Fuzzy System (NFS), a hybrid intelligent
	system, which combines Fuzzy Logic and Artificial Neural Network.
	The fuzzy systems capable to mimic human-like reasoning through the
	use of the fuzzy sets and fuzzy linguistic rule, whereas the neural network plays important role as a basis
	to develop algorithm to model complex patterns and prediction problems.
	
	In the real world applications, systems are usually assembled with
	shifts and drifts, which cannot be handled by the common NFSs. These NFSs usually really depend on expert knowledge \cite{tung2010efsm} and in turn requires
	tedious manual interventions. This problem leads to a static rule
	base, which unable to be adjusted once initial setting is setup to
	gain a better performance. PANFIS is built to further develop the
	disadvantage of common classical NFS to overcome the degree of nonlinearity
	which exists in the real world systems. 
	
	PANFIS is able to assemble a complex system's rule base autonomously
	by learning system from scratch with an empty rule base. Fuzzy rules
	and its parameters can be generated, updated and pruned during the learning
	learning process. Furthermore, the rule merging process is also
	executed during the learning by identifying identical fuzzy rule sets
	to simplify the rules complexity.
	
	The main characteristics of PANFIS is the generation of ellipsoids
	in arbitrary direction. These ellipsoids describe the local correlation
	between variables. The antecedent parts of fuzzy sets are formed by the ellipsoids connected with a new projection. 
	Thus, it is more interpretable for an expert/user in comparison with the non-axis parallel model applied in \cite{lemos2011multivariable}.
	
	PANFIS system evolution is initialized by the generation of rules which
	is driven by datum significance (DS) criterion. DS algorithm was proposed
	in \cite{huang2005generalized} and \cite{rong2006sequential} to
	identify the datum's high-potential managing troublesome data streams.
	This initialization of rule growing must assures that $\varepsilon$-completeness
	criterion to ensure that rule base and the fuzzy partitions have a
	sufficient coverage of the input space. During the system's evolution,
	when the new datum is covered by the current rules, the rule parameters
	(focal points and radii) of the fuzzy rules and fuzzy consequences
	are updated based on the new datum attributes. For focal points and
	radii, original PANFIS applies Extended Self-organizing Map (ESOM)
	theory by adjusting neighboring rules. However, in this work, rule
	adaptation utilizes GENEFIS \cite{pratama2014genefis}, pClass \cite{pratama2015pclass},
	and GEN-SMART-EFS \cite{lughofer2015generalized}. This is due the
	ESOM's drawbacks in regards to the possibility of instability issues
	when the inverse covariance matrix is ill-conditioned (e.g., due to
	redundant input features). Fuzzy consequences are also adjusted by
	using the enhanced recursive least square (ERLS), an extension of
	conventional recursive least square (RLS). In the ENFS community,
	ERLS has become the common method which is proven to support system
	errors' convergence and the update of weight vector.
	
	The rule pruning of PANFIS is inherited from rule base simplification
	technology namely generalized growing and pruning radial basis function
	(GGAP-RBF) in SAFIS\cite{rong2006sequential}. However, rule base
	simplification in SAFIS cannot be applied in PANFIS as it is only suits
	to zero-order TSK fuzzy system and the unidimensional membership function
	environments. PANFIS extends this method using extended rule significance
	(ERS) concept by defining the rule significant as statistical contribution
	of the fuzzy rule when the number of observation approaches to infinity.
	Furthermore, the statistical contribution of the fuzzy rule over all
	fuzzy rules is determined by the fuzzy rule volume over the all rules
	volume. If the contribution vector value of such rules less than the threshold $kerr$, these rules are regarded as outdated rules and will be discarded to
	reduce the network complexity.
	
	PANFIS also applies fuzzy set merging process to form effective (economical
	and interpretable) rule base, and also to achieve a good predictive
	quality. This is done due to the identical/overlapping fuzzy sets
	(e.g., due to their similarity in membership functions). This similarity
	can be measured by benefiting a kernel-based metric method comparing
	the center and the widths of two fuzzy sets in one joint formula \cite{lughofer2011line}.
	In the case of the perfect match, the two fuzzy sets are identical,
	and has the degree of similarity measure equal to 1. Conversely, if
	the two fuzzy sets have the similarity degree above the tolerable
	value $Sker \geqslant 0.8$ based on \cite{lughofer2011line}, the two fuzzy
	sets could be merged.
	
	\subsection{Big Data Analytic Architecture and Data Flow Process in Spark Platform\label{subsec:BigDataFlow}}
	
	Apache Spark (Spark) is regarded as the latest framework for big data
	analytic to support advanced in-memory programming model. Spark ecosystem
	consists of two main parts: lower-level libraries known as Spark-core
	and upper-level libraries known as extension of Spark-core. Spark-core
	consists of some programming interfaces such as Scala, Java, Python,
	R, and SQL as an integrated Spark APIs. Upper-level libraries
	consists of Spark's MLib for machine learning purposes, GraphX for
	graph analysis, Spark Streaming for stream processing, and Spark SQL
	for structured data processing. All of these Spark's components enable
	Spark to perform scalable distributed machine learning, graph analysis,
	and structured data processing.
	
\begin{figure}[htbp]
		\begin{centering}
			\includegraphics[width=9cm]{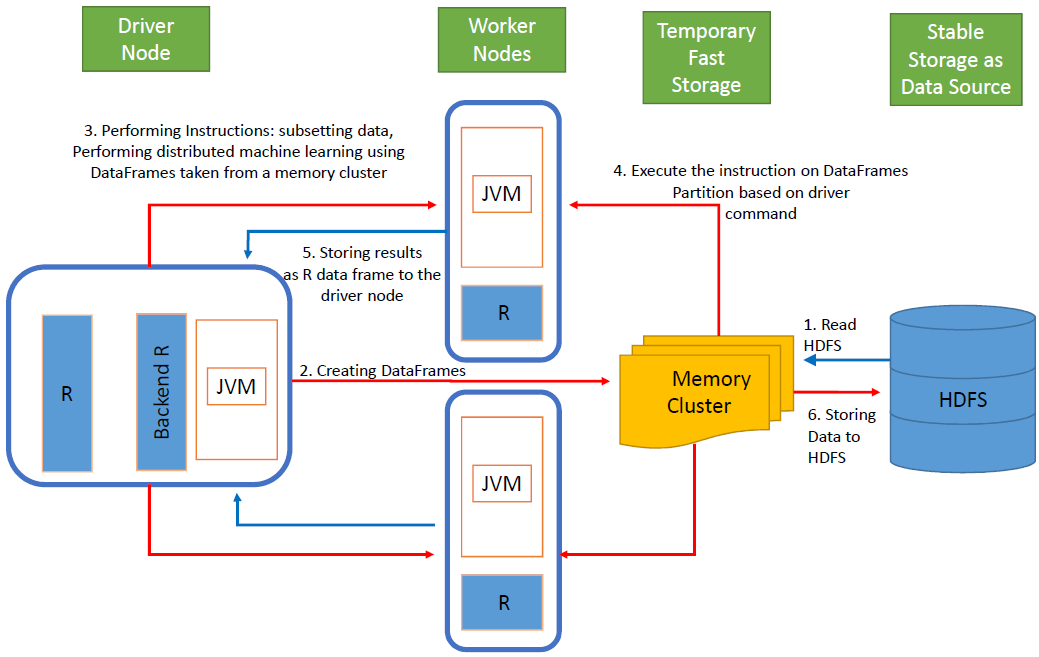}
			\par\end{centering}
		\caption{Data control flow and operation in Distributed Machine Learning based on PANFIS\label{Figure1}}
	\end{figure}

	In this work, we utilize SparkR, an R package distributed with Spark to provide R front-end to Spark. 
	It enables R to manipulate Spark DataFrames (DataFrames), a Spark data abstraction that is fault-tolerant for in-memory cluster computing, and operates Spark functionality in R as shown in Fig. \ref{Figure1}. Fig. \ref{Figure1}  shows the control flow of distributed big data analytics in this framework. 
	\begin{figure}[htbp]
		\begin{centering}
			\includegraphics[width=9cm]{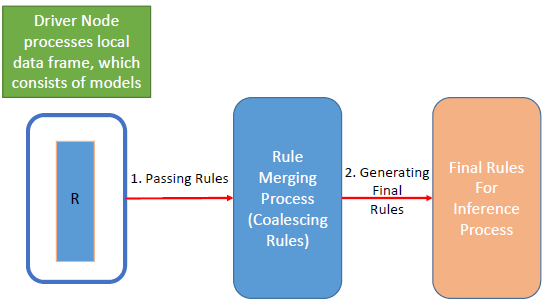}
			\par\end{centering}
		\caption{The Rule Merging Scheme of processing all models into a single model in local R data frame\label{Figure2}}
	\end{figure}
	
	The big data analytic framework consists of nodes (driver node and some worker nodes) and the data sources (e.g, Hadoop Distributed File System (HDFS)). In this framework, distributed machine learning process is initialized by loading RFID dataset from HDFS storage, which is shown at step 1 Fig. \ref{Figure1}. The second step, the driver sends an instruction to create the DataFrames (Spark's data abstraction) then stores the DataFrames in a memory cluster, a shared memory in the cluster. DataFrames is the only data type that can be distributed in the Spark platform. At the third step, the driver node sends many instructions to manipulate the DataFrames. These instructions can be repetitively created by the driver node to manipulate DataFrames, for example, performing distributed algorithm (PANFIS) to all worker nodes to learn the chunk of DataFrames. Worker will process a chunk with a function (e.g. PANFIS) applied to them and send the result (rule(s)/model) to the driver node, which is shown at step 4 Fig. \ref{Figure1}. At fifth step, all models generated from the worker nodes will be converted into local driver node memory as R data frame as a collected models, which can be seen by user. Step 6, is an optional step, whether driver node want to stores all models in the HDFS. Please be noted that all the instructions created by the R front-end is translated by JVM to be able to access Spark API to manipulate the DataFrames.

	As explained in Fig. \ref{Figure1}, driver node receives all rules (models) generated from many worker nodes when they process all data chunks using PANFIS algorithm. Fig. \ref{Figure2} depicts the data frame of R is captured in driver node as a learning results (a collected models). These results (rules) are then merged become the final model for inference purpose based on the process described in subsection \ref{subsec:RuleMerging}.

	\subsection{Rule Merging of Final Rules after All Data Chunk's Learning \label{subsec:RuleMerging}}
	
	Distributed big data analytic generates rules for every data chunk
	processed in the processors/nodes. In the distributed big data analytic
	architecture, processors/nodes are embedded by machine learning algorithm,
	which is in this case PANFIS algorithm. Rules generated by PANFIS
	represent the clusters of the data chunk. With so many rules generated
	in the system, some rules may become overlapping. Some rules can be
	merged to reduce the redundancy between the rules \cite{lughofer2011evolving}.
	This merging process is essential in order to ensure the readability
	and transparency and avoiding the ambiguous rules. This rule merging
	scenario refers to the explanation in \cite{lughofer2015generalized} without similarity criterion to extend the PANFIS rule merging scenario. The implementation of rule-merging
	follows two criterions: 1) Inspecting the 2 rules to be merged
	whether they are lying nearby, touching, or even slightly overlapping;
	2) Calculating the homogeneity of the adjacent rules. 
	
	The first criterion in merging process is calculating the degree of
	overlapping for two clusters which measured by Bhattacharrya distance
	formula \cite{bhattacharyya1943measure}\cite{djouadi1990quality}.
	This criterion process is calculated after the rules collection process
	in the learning phase. The overlap degree between target cluster (rule)
	expressed by \emph{win }and the other clusters, \emph{$k=\{1,...,C\}\text{\textbackslash}\{win\}$}
	is expressed as follow :
	
	\begin{equation}
	\begin{array}{c}
	olap(win,k)=\frac{1}{8}(c_{win}-c_{k})^{T}\sum^{-1}(c_{win}-c_{k})\\
	+\frac{1}{2}\ln(\frac{\det\sum^{-1}}{\sqrt{\det\sum_{win}^{-1}\det\sum_{k}^{-1}}})
	\end{array}\label{FormulaOlap(win,k)}
	\end{equation}
	
	where\emph{ $\Sigma^{-1}=(\Sigma_{win}^{-1}+\Sigma_{k}^{-1})/2$.
	}The highest value of statistical contribution is chosen among the
	clusters, which based on the equation explained in PANFIS \cite{pratama2014panfis} as the target cluster \emph{win}. The distance between two ellipsoidal
	clusters has the value of 0 if they both are touching, greater than
	0 in the case of overlapping, and less than zero for disjoint situation.
	The minimum threshold for clusters to be merged is 0, which is the
	condition where both clusters are touching each other in regard to
	the conditions stated.
	
	The homogeneity of adjacent rules is the second criterion of the merging process. This criterion ensures the homogeneous shape and direction of two merged rules. This criterion also reflects the two local data clouds. By utilizing the blow up effect to trigger homogeneous joint regions, the homogeneity is defined as follows: 
	
	\begin{equation}
	V_{merged}\leqslant p(V_{win}+V_{k})\label{eq:FormulaVmerged}
	\end{equation}
	\label{mergevolume}where $p$ is input attribute's dimension.
	$V_{merged}$, $V_{win}$, and $V_{k}$ depict merged
	rules' volume, winning rule, and compared rule respectively. Rule's volume can refer to the formula explained in PANFIS \cite{pratama2014panfis}.
	In conclusion, the merging criterions can be simplified by the conditions as follow:
	
	\begin{equation}
	(Eq.(\ref{FormulaOlap(win,k)})\geq thr)|Eq.(\ref{eq:FormulaVmerged})\label{eq:Formula Gabung}
	\end{equation}
	\label{allcriterions}where $thr$ is set to 0.8 based on empirical
	experience in \cite{lughofer2011line}.
	
	\subsubsection{Rule merging phase's policy\label{subsec:Rule-merging-policy second phase}}
	
	Merging process will be conducted if two criterions are satisfied refer to the winning rule. The winning rule refers to the rule which has more data points than another rule $N_{win}>N_k$. 
	The updating parameters of rule merging is executed based on the work in  \cite{pratama2014genefis} which is formulated as follows:
	
	\begin{equation}
	\begin{array}{c}
	c_{win}^{new}=\frac{c_{win}^{old}N_{win}^{old}+c_{k}^{old}N_{k}^{old}}{N_{win}^{old}+N_{k}^{old}}
	\end{array}\label{newcenter}
	\end{equation}
	
	\begin{equation}
	\begin{array}{c}
	\sum_{win}^{-1}(new)=\frac{\sum_{win}^{-1}(old)*N_{win}^{old}+\sum_{k}^{-1}(old)*N_{k}^{old}}{N_{win}^{old}+N_{k}^{old}}
	\end{array}\label{newspread}
	\end{equation}
	
	\begin{equation}
	\begin{array}{c}
	N_{win}^{new}=N_{win}^{old}+N_{k}^{old}
	\end{array}\label{newpop}
	\end{equation}
	
	\begin{equation}
	\begin{array}{c}
	w_{win}^{new}=\frac{w_{win}^{old}*N_{win}^{old}+w_{k}^{old}*N_{k}^{old}}{N_{win}^{old}+N_{k}^{old}}
	\end{array}\label{newweight}
	\end{equation}
	
	Where $c_{win}^{new}$ and $\sum_{win}^{-1}(new)$ are the new antecedent parameters of the merged rule and $w_{win}^{new}$ is the consequent parameter of the merged rule.
	
	\section{Experimental setup and results}
	
	The environmental setup in which Scalable PANFIS is applied in the distributed framework along with the experiment results will be described in this section. The framework performance between PANFIS distributed framework (Scalable PANFIS), PANFIS, and some other machine learning distributed frameworks are compared in order to measure the running time and accuracy. The data used in this experiment is RFID data, generated from real application based on sensor in classification problem. This dataset consists of 283,100 instances with only 1 dimension of input describing the RSS frequency of RFID to infer the location of Manufacturing Object (MOs).  
	
	\subsection{Experiments\label{subsec:experiments}}
	NeCTAR Cloud, which provides flexible scalable computing is used as the computing environment for this experiment. 
	The framework consists of 7 nodes (1 driver and 6 workers). Each node has the specification as following: 30 GB Disk Capacity, 6GB RAM, and NeCTAR Ubuntu 16.04 LTS (Xenial) amd64 as operating system. The total cluster memory used in this framework is 30 GB as we use 5 GB of memory for 6 worker nodes, leaving another 1 GB for every node for other processes in the nodes. We use spark 2.2.1 release for cluster computing system at the time of writing. In this experiment 5 algorithms are compared: Scalable PANFIS, PANFIS, generalize linear regression model (GLM), Gradient Boost Tree (GBT) and KMeans. 
	For comparison purpose with other seminal evolving algorithms, we also conduct the experiment against gClass \cite{pratama2016incremental}, eT2Class \cite{pratama2016evolving}, pClass \cite{pratama2015pclass}, and eT2ELM \cite{pratama2016incremental}. For this purpose, we use the the personal computer with the following specifications: Intel(R) Core(TM) i7-6700 CPU @3.4 GHz 16GB RAM and Matlab 2017b Software specification.
	
	These algorithms are chosen to perform the experiments based on the following reasons:
	\begin{enumerate}
		\item PANFIS is a seminal evolving algorithm which has capability to learn data stream in online 
		manner. In this case, this base algorithm is compared against the proposed method, big data analytic based on PANFIS (Scalable PANFIS).
		
		\item Random Forest (RF), Generalized Linear Model (GLM), Gradient Boost Tree (GBT), KMeans, are chosen as part of Machine Learning Library (MLib) on sparkR as a scalable machine learning for knowledge discovery. Thus we use these algorithms to compare their performance against scalable PANFIS. However, as their nature are binary classifier algorithm, we apply one vs rest strategy to be able to act as multi-class classifier.
		\item gClass, eT2Class, pClass, and eT2ELM are compared in terms of running time and accuracy to perform the classification task as they are also classified as a seminal evolving algorithms.
		
	\end{enumerate}
	
	The classification scenarios will measure the running time and the
	accuracy to evaluate the performance of all algorithms. The details of the experiment results are explained in the subsection \ref{subsec:result}.
	
	\subsection{Results\label{subsec:result}}
	This subsection details the performance of every classifier in classifying the RFID dataset as described in subsection \ref{subsec:rfidapps}. The dataset is divided into two parts: 1) 200,000 of training data and 2) 83,100 of test data for validation. We conduct the experiment for five different algorithms: The first algorithm, PANFIS, is executed in single  CPU without distributed processing with the specification of 30 GB Disk Capacity, 6GB RAM, and NeCTAR Ubuntu 16.04 LTS (Xenial) amd64 as operating system. The software used for running PANFIS is R version 3.4.3. The second to fifth algorithm are the Scalable PANFIS and four other MLib algorithms provided in the Spark library, which are executed with distributed processing in the Spark platform. We utilize SparkR to provide a lightweight front end for using Apache Spark from R. The number of data partition used for experiment in the distributed environment system is 50 partitions.
	
	The result depicted in Table \ref{tab:Tabel1} shows that PANFIS algorithm (single machine learning) performs significant result with 98.71 percent of accuracy in learning the RFID data. The Scalable PANFIS, a distributed big data analytic based on PANFIS machine learning in the Spark environment, yields 96.67 percent in accuracy, which is still comparable with PANFIS single machine learning. The merging process as shown in Table \ref{tab:Tabel2}, reduces the number of rules from 55 into around half of 28 rules. The comparison
	of Scalable PANFIS with three other algorithms provided
	in MLib shows that Scalable PANFIS can handle the nonstationary
	data stream with the accuracy higher than GLM, GBT, and Kmeans with 96.67, 50.03,
	75.07, and 47.66 percent for Scalable PANFIS, GLM, GBT,
	and KMeans respectively. Scalable PANFIS is slightly outperformed
	by RF with 96.67 and 98.56 percent for Scalable
	PANFIS and RF respectively. However, Scalable PANFIS
	better than RF in terms of running time. Please be noted that
	all of the MLib algorithms are binary classifier in nature and
	are converted as multi-class classifier by applying one vs rest
	strategy.
	
	Another highlight is also depicted in the Table \ref{tab:Tabel1} in terms of running time. Scalable PANFIS performs very significant result with more than 20 times faster than PANFIS executed in the single CPU with 104 seconds and 2130 seconds for Scalable PANFIS and PANFIS respectively in terms of running time.

	To further evaluate the Scalable PANFIS performance, this
	algorithm is benchmarked with other state-of-the-art algorithms:
	eT2Class, eT2ELM, gClass, and pClass. Table \ref{tab:Tabel3} shows that the accuracy and the running time of other
	benchmarked algorithms. This experiment is conducted in
	the same computational environment. The result shows that
	Scalable PANFIS has a comparable performance in terms of
	accuracy with other evolving algorithms.

	\begin{table}[htbp]
		\caption{The performance comparison of Scalable PANFIS against single PANFIS and other MLibs algorithms}\label{tab:Tabel1}
		\begin{center}
			\begin{tabular}{|c|c|c|}
				\hline 
				{\textbf {Algorithm} }& {\textbf {Accuracy (\%)}} & {\textbf {Running Time (s)}}\tabularnewline
				
				\hline 
				PANFIS & 98.71 & 2130\tabularnewline
				\hline 
				Scalable PANFIS & 96.67 & 104\tabularnewline
				\hline 
				RF & 98.56 & 149\tabularnewline
				\hline 
				GLM  & 50.03 & 264\tabularnewline
				\hline 
				GBT & 75.07 & 272\tabularnewline
				\hline 
				Kmeans & 47.66 & 88\tabularnewline
				\hline 
			\end{tabular}
		\end{center}
	\end{table}
	\begin{table}[htbp]
		\caption{The Evolution Number of Rules in the Merging Rule process of big data analytic framework }\label{tab:Tabel2}
		\begin{center}
			
			\begin{tabular}{|c|c|c|c|}
				\hline 
				\multicolumn{2}{|c|}{\textbf{Number of Rule} } \\ 
				\hline 
				Before Merging & After Merging \\ 
				\hline 
				55 & 28 \\ 
				\hline 
			\end{tabular} 
			
			\label{tab1}
		\end{center}
	\end{table}
	\begin{table}[htbp]
		\caption{Performance of other seminal evolving algorithms running standalone in the single CPU using matlab environment}\label{tab:Tabel3}
		\begin{center}
			\begin{tabular}{|c|c|c|c|}
				
				\hline
				{\textbf{Algorithm}}& {\textbf{Accuracy (\%)}}&{\textbf{Time (s)}}\\
				
				\hline
				eT2Class& 98.77&1223  \\
				\hline
				eT2ELM& 95.38&350  \\
				\hline
				gClass& 92.89&1253  \\
				\hline
				pClass& 98.05&604  \\
				\hline
			\end{tabular}
			\label{tab1}
		\end{center}
	\end{table}

	\section{Conclusion and Future Works}
	Scalable PANFIS is a big data analytic framework which processes high-volume of big data by distributing data stream into many partitions thus accelerate the learning process. PANFIS algorithm, the base algorithm of Scalable PANFIS learns the data chunk in online manner. The models generated from the learning process of many data partitions can be merged become a single model without reducing the accuracy performance.
	For the future work, we will further evaluate the performance of Scalable PANFIS by testing this algorithm to many other high-dimensional big data with using some classification techniques.
	
	\section{Acknowledgment}
	This project is fully supported by NTU start up grant and MOE tier 1 research grant. This research is also supported by use of the Nectar Research Cloud, a collaborative Australian research platform supported by the National Collaborative Research Infrastructure Strategy (NCRIS).
	
	\bibliographystyle{IEEEtran}
	\bibliography{IEEEabrv,IEEEexample,biblatex-examples,UpdateBib0502}

\begin{thebibliography}{10}
\providecommand{\url}[1]{#1}
\csname url@samestyle\endcsname
\providecommand{\newblock}{\relax}
\providecommand{\bibinfo}[2]{#2}
\providecommand{\BIBentrySTDinterwordspacing}{\spaceskip=0pt\relax}
\providecommand{\BIBentryALTinterwordstretchfactor}{4}
\providecommand{\BIBentryALTinterwordspacing}{\spaceskip=\fontdimen2\font plus
\BIBentryALTinterwordstretchfactor\fontdimen3\font minus
  \fontdimen4\font\relax}
\providecommand{\BIBforeignlanguage}[2]{{%
\expandafter\ifx\csname l@#1\endcsname\relax
\typeout{** WARNING: IEEEtran.bst: No hyphenation pattern has been}%
\typeout{** loaded for the language `#1'. Using the pattern for}%
\typeout{** the default language instead.}%
\else
\language=\csname l@#1\endcsname
\fi
#2}}
\providecommand{\BIBdecl}{\relax}
\BIBdecl

\bibitem{ni2011rfid}
L.~M. Ni, D.~Zhang, and M.~R. Souryal, ``Rfid-based localization and tracking
  technologies,'' \emph{IEEE Wireless Communications}, vol.~18, no.~2, 2011.

\bibitem{za2017evolving}
C.~Za'in, M.~Pratama, E.~Lughofer, and S.~G. Anavatti, ``Evolving type-2 web
  news mining,'' \emph{Applied Soft Computing}, vol.~54, pp. 200--220, 2017.

\bibitem{pratama2016evolving}
M.~Pratama, J.~Lu, and G.~Zhang, ``Evolving type-2 fuzzy classifier,''
  \emph{IEEE Transactions on Fuzzy Systems}, vol.~24, no.~3, pp. 574--589,
  2016.

\bibitem{pratama2015pclass}
M.~Pratama, S.~G. Anavatti, M.~Joo, and E.~D. Lughofer, ``pclass: an effective
  classifier for streaming examples,'' \emph{IEEE Transactions on Fuzzy
  Systems}, vol.~23, no.~2, pp. 369--386, 2015.

\bibitem{lughofer2015generalized}
E.~Lughofer, C.~Cernuda, S.~Kindermann, and M.~Pratama, ``Generalized smart
  evolving fuzzy systems,'' \emph{Evolving Systems}, vol.~6, no.~4, pp.
  269--292, 2015.

\bibitem{sayed2012learning}
M.~Sayed-Mouchaweh and E.~Lughofer, \emph{Learning in non-stationary
  environments: methods and applications}.\hskip 1em plus 0.5em minus
  0.4em\relax Springer Science \& Business Media, 2012.

\bibitem{shvachko2010hadoop}
K.~Shvachko, H.~Kuang, S.~Radia, and R.~Chansler, ``The hadoop distributed file
  system,'' in \emph{Mass storage systems and technologies (MSST), 2010 IEEE
  26th symposium on}.\hskip 1em plus 0.5em minus 0.4em\relax Ieee, 2010, pp.
  1--10.

\bibitem{zaharia2010spark}
M.~Zaharia, M.~Chowdhury, M.~J. Franklin, S.~Shenker, and I.~Stoica, ``Spark:
  Cluster computing with working sets.'' \emph{HotCloud}, vol.~10, no. 10-10,
  p.~95, 2010.

\bibitem{fernandez2014big}
A.~Fern{\'a}ndez, S.~del R{\'\i}o, V.~L{\'o}pez, A.~Bawakid, M.~J. del Jesus,
  J.~M. Ben{\'\i}tez, and F.~Herrera, ``Big data with cloud computing: an
  insight on the computing environment, mapreduce, and programming
  frameworks,'' \emph{Wiley Interdisciplinary Reviews: Data Mining and
  Knowledge Discovery}, vol.~4, no.~5, pp. 380--409, 2014.

\bibitem{zabig}
C.~Zain, M.~Pratama, E.~Lughofer, M.~M. Ferdaus, Q.~Cai, and M.~Prasad, ``Big
  data analytics based on panfis mapreduce.''

\bibitem{pratama2014panfis}
M.~Pratama, S.~G. Anavatti, P.~P. Angelov, and E.~Lughofer, ``Panfis: A novel
  incremental learning machine,'' \emph{IEEE Transactions on Neural Networks
  and Learning Systems}, vol.~25, no.~1, pp. 55--68, 2014.

\bibitem{tung2010efsm}
W.~L. Tung and C.~Quek, ``efsm-a novel online neural-fuzzy semantic memory
  model,'' \emph{IEEE Transactions on Neural Networks}, vol.~21, no.~1, pp.
  136--157, 2010.

\bibitem{lemos2011multivariable}
A.~Lemos, W.~Caminhas, and F.~Gomide, ``Multivariable gaussian evolving fuzzy
  modeling system,'' \emph{IEEE Transactions on Fuzzy Systems}, vol.~19, no.~1,
  pp. 91--104, 2011.

\bibitem{huang2005generalized}
G.-B. Huang, P.~Saratchandran, and N.~Sundararajan, ``A generalized growing and
  pruning rbf (ggap-rbf) neural network for function approximation,''
  \emph{IEEE Transactions on Neural Networks}, vol.~16, no.~1, pp. 57--67,
  2005.

\bibitem{rong2006sequential}
H.-J. Rong, N.~Sundararajan, G.-B. Huang, and P.~Saratchandran, ``Sequential
  adaptive fuzzy inference system (safis) for nonlinear system identification
  and prediction,'' \emph{Fuzzy Sets and Systems}, vol. 157, no.~9, pp.
  1260--1275, 2006.

\bibitem{pratama2014genefis}
M.~Pratama, S.~G. Anavatti, and E.~Lughofer, ``Genefis: toward an effective
  localist network,'' \emph{IEEE Transactions on Fuzzy Systems}, vol.~22,
  no.~3, pp. 547--562, 2014.

\bibitem{lughofer2011line}
E.~Lughofer, J.-L. Bouchot, and A.~Shaker, ``On-line elimination of local
  redundancies in evolving fuzzy systems,'' \emph{Evolving Systems}, vol.~2,
  no.~3, pp. 165--187, 2011.

\bibitem{lughofer2011evolving}
E.~Lughofer, \emph{Evolving fuzzy systems-methodologies, advanced concepts and
  applications}.\hskip 1em plus 0.5em minus 0.4em\relax Springer, 2011,
  vol.~53.

\bibitem{bhattacharyya1943measure}
A.~Bhattacharyya, ``On a measure of divergence between two statistical
  populations defined by their probability distribution,'' \emph{Bull. Calcutta
  Math. Soc}, 1943.

\bibitem{djouadi1990quality}
A.~Djouadi, O.~Snorrason, and F.~Garber, ``The quality of training sample
  estimates of the bhattacharyya coefficient,'' \emph{IEEE Transactions on
  Pattern Analysis and Machine Intelligence}, vol.~12, no.~1, pp. 92--97, 1990.

\bibitem{pratama2016incremental}
M.~Pratama, J.~Lu, S.~Anavatti, E.~Lughofer, and C.-P. Lim, ``An incremental
  meta-cognitive-based scaffolding fuzzy neural network,''
  \emph{Neurocomputing}, vol. 171, pp. 89--105, 2016.

\end{thebibliography}

\end{document}